\documentclass[10pt]{article}

\usepackage{amsmath,amsfonts}
\usepackage{mathtools} 
\usepackage[cbgreek]{textgreek}
\usepackage{bm}
\usepackage[colorlinks=true, allcolors=blue]{hyperref}
\usepackage{cite}

\renewcommand{\sp}{\hspace{1pt}}
\newcommand{\nsp}{\hspace{-1pt}}

\newcommand{\ui}{\mathrm{i}}
\newcommand{\ud}{\mathrm{d}}
\newcommand{\bu}{{\bf u}}
\newcommand{\bx}{{\bf x}}
\newcommand{\bp}{{\bf p}}
\newcommand{\bq}{{\bf q}}
\newcommand{\bs}{{\bf s}}

\newcommand{\JonesMat}[5][-4]{\biggl[\hspace{-2pt}
	\begin{array}{cc}
		#2 & #3 \\[-0 pt]
		#4 & #5
	\end{array}
	\hspace{-2pt}
	\biggr]
}

\DeclarePairedDelimiter{\norm}{\lVert}{\rVert}
\DeclarePairedDelimiter{\abs}{\lvert}{\rvert}

\newcommand{\VellaChange}[1]{#1} 

\usepackage{placeins}
\usepackage{authblk}

\title{Poincar\'e sphere representation for spatially varying birefringence}
\author[*]{Anthony Vella}
\author[ ]{Miguel A. Alonso}

\affil[ ]{\!\!\normalsize The Institute of Optics, University of Rochester, Rochester NY 14627, USA}
\affil[*]{\normalsize Corresponding author: avella@optics.rochester.edu}

\begin{document}
	
\maketitle

\begin{center}
	{\bf Abstract}
\end{center}
\vspace{-5pt}
The Poincar\'e sphere is a graphical representation in a three-dimensional space for the polarization of light. Similarly, an optical element with spatially varying birefringence can be represented by a surface on a four-dimensional ``Poincar\'e hypersphere''. A projection of this surface onto the traditional Poincar\'e sphere provides an intuitive geometric description of the polarization transformation performed by the element, as well as the induced geometric phase. We apply this formalism to quantify the effects of birefringence on the image quality of an optical system.

\section{Introduction}
Several recent technologies have enabled the production of optical elements with tailored spatially varying birefringence, allowing the generation of beams with complex polarization patterns \cite{Hasman_2005}. These technologies include metasurfaces composed of plasmonic \cite{Kildishev_2013,
Yu_2014} or dielectric \cite{Bomzon_2002,Arbabi_2015,Kruk_2016,Genevet_2017,Devlin_2017} nanostructures, as well as liquid crystal devices such as $q$-plates \cite{Marrucci_2006,McEldowney2008_vortex,
Marrucci_2013}, light valves \cite{Aubourg_1982,Aleksanyan_2017}, and spatial light modulators \cite{Moreno_2012}.
Spatially varying birefringence also occurs naturally in standard materials like plastic or glass, due to internal mechanical stress. Stress-induced birefringence in optical elements can result from their manufacture process or be caused by their mount, and often has undesirable effects on their optical performance \cite{Doyle_2002}. It is worth mentioning, though, that stress 
can also be tailored to produce birefringence distributions \cite{Spilman_2007_vortex} that are useful in polarimetry \cite{Ramkhalawon_2013,Zimmerman_2016,Sivankutty_2016} or for the generation of beams with interesting polarizations \cite{Beckley_2010,Vella_2017_bottle,Vella_2017_bottle_erratum}.

In this work we propose a geometric description of spatially varying birefringence distributions, whether they are designed or accidental, based on a generalization of the Poincar\'e sphere, which is usually employed to describe beams and not materials. Our assumption is that the material is transparent (i.e. absorption is negligible), static (i.e. it induces no depolarization), and thin (so that the polarization transformation it induces is local).

\section{Jones matrix of a birefringent mask}\label{sect:General_BM}
The Jones matrix of a thin, transparent birefringent mask (BM) can in general be written as \cite{Vella_2017_bottle}
\begin{equation}
\mathbb{J}(\bx)=\exp(\ui\Gamma)\bigl[{\bf p}_1{\bf p}_1^\dagger \exp(-\ui\delta)
+ \,{\bf p}_2{\bf p}_2^\dagger\exp(\ui\delta)\bigr],\label{eq:JonesBM_general}
\end{equation}
where $\bx$ is a two-dimensional spatial coordinate determining a position over the surface of the mask, $\Gamma(\bx)$ is a global phase function, $\bp_{1,2}(\bx)$ are the two (not necessarily linear) eigenpolarizations at each point of the BM, $\delta(\bx)$ is half the phase mismatch (retardance) between these eigenpolarizations, and $\bp_j^\dagger$ is a conjugate transpose. Since $\mathbb{J}$ is invariant to a full-wave retardance increment, we may restrict $\delta\in[0,\pi]$ without loss of generality. In fact, any BM can be represented within the range $\delta\in[0,\pi/2]$, since the substitution $\delta\rightarrow\pi-\delta$ simply reverses the roles of $\bp_1$ and $\bp_2$ and introduces a $\pi$ phase shift that can be absorbed by $\Gamma$. 

We assume a transparent mask, so the Jones matrix is unitary and hence $\Gamma$ and $\delta$ are real and the eigenpolarizations ${\bf p}_j$ are orthonormal. For convenience, in what follows we use the circular polarization basis. The eigenpolarizations of the BM are
\begin{equation}
\bp_{1,2}(\bx) = \frac{1}{\sqrt{2}}
\left[\!\!\begin{array}{l}
\left[\pm\cos(\Theta/2) + \sin(\Theta/2)\right]e^{-\ui\Phi/2} \\[2pt]
\phantom{\pm}\left[\cos(\Theta/2) \mp \sin(\Theta/2)\right]e^{\ui\Phi/2}
\end{array}\!\!\right],
\end{equation}
where the functions $\Theta(\bx)\!\in\![-\pi/2,\pi/2]$ and $\Phi(\bx)\!\in\![0,2\pi]$ are the latitude and longitude angles of ${\bf p}_1(\bx)$ over the Poincar\'e sphere. By expanding Eq.~(\ref{eq:JonesBM_general}), the Jones matrix may be written as
\begin{equation}
\mathbb{J}(\bx) = \exp(\ui\Gamma)\JonesMat{q_0-\ui q_3\;}{-q_2-\ui q_1}{q_2-\ui q_1\;}{\phantom{-}q_0+\ui q_3},
\end{equation}
where the $q_n$ are the elements of a unit four-vector $\vec{q}(\bx)$, given \nolinebreak[4] by%
\begin{subequations}\label{eq:q_hyper2}
	\begin{align}
	q_0(\bx) &= \cos\delta,\\
	q_1(\bx) &= \sin\delta\cos\Theta\cos\Phi,\\
	q_2(\bx) &= \sin\delta\cos\Theta\sin\Phi,\\
	q_3(\bx) &= \sin\delta\sin\Theta.
	\end{align}
\end{subequations}

\section{Poincar\'e sphere and hypersphere}
We now provide a geometric description of the birefringence distribution over the same three-dimensional Poincar\'e sphere that describes the field's polarization. The polarization state of the incident field ${\bf E}_0$ can be represented on the Poincar\'e sphere by its normalized Stokes vector $\bs=(s_1,s_2,s_3)$.
At each point over the BM, the local eigenpolarizations $\bp_{1,2}$ have Stokes parameters $\bs_{\bp_{1,2}}=\pm(\cos\Theta\cos\Phi,\cos\Theta\sin\Phi,\sin\Theta)$, which correspond to antipodal points on the surface of the Poincar\'e sphere. 


Similarly, the unit vector $\vec{q}$ is constrained to the hypersurface of a 4D unit hypersphere (the ``Poincar\'e hypersphere''), described by a polar angle $\delta$ and the latitude and longitude angles $\Theta$ and $\Phi$. As mentioned earlier, replacing $\delta\to\pi-\delta$ is equivalent to swapping $\bp_1$ and $\bp_2$, that is, to changing $\Theta\to-\Theta$ and $\Phi\to\Phi+\pi$. Any pair of antipodal points $\vec{q}$ and $-\vec{q}$ on the Poincar\'e hypersphere then correspond to the same birefringence
, so only the upper half of the hypersphere (where $\delta\in[0,\pi/2]$, $q_0\ge0$) is needed to describe an arbitrary BM.
This half of the hypersphere can be projected onto the solid 3D Poincar\'e sphere by dropping the coordinate $\smash{q_0=(1-\abs{\bq}^2)^{1/2}}$, 
where $\bq=(q_1,q_2,q_3)$. The resulting projection $\bq$ onto the Poincar\'e sphere lies in the direction of $\bs_{\bp_1}$ at a distance $\abs{\bq}=\sin\delta$ from the origin. 
Note that points for which $\delta\in(\pi/2,\pi]$ (that is, $q_0<0$) must be mapped onto $-\bq$, and therefore a smooth transition in which $\delta$ crosses $\pi/2$ corresponds to points leaving the 3D Poincar\'e sphere at one point over its surface and reentering at the opposite point (with a $\pi$ phase offset, as discussed above).

\begin{figure}[htb]
	\centering
	\includegraphics[width=.8\linewidth]{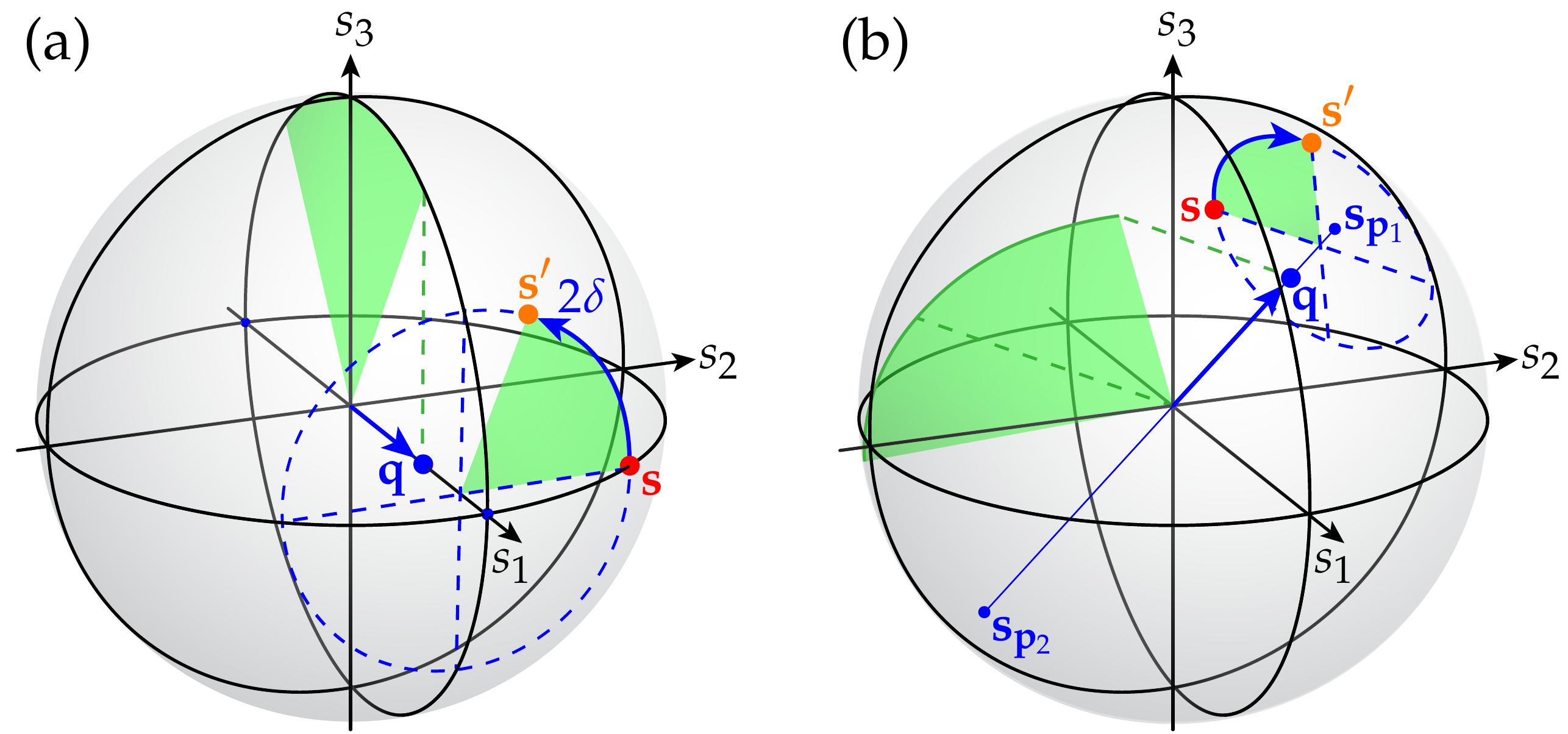}
	\caption{Rotation of the input polarization at a single point on the BM, illustrated for cases where the input polarization and local BM eigenpolarizations and are (a) linear and (b) elliptical. Since $\abs{\bq}\nsp=\sin\delta$, an orthogonal projection of $\bq$ onto the surface of the sphere spans half the rotation angle (shown in green).}
	\label{fig:PoincareRotation}
\end{figure}

At each point of the BM, the local effect on the incident polarization $\bs$ is a rotation on the Poincar\'e sphere about the axis $\overline{\bs_{\bp_1}\bs_{\bp_2}}$ through an angle $2\delta$ \cite{Baylis_1993,Ossikovski_2013}. The quaternion algebra formalism, which has been used previously to describe polarized light \cite{Richartz_1949, Pellat_1990, Pellat_1991,Ainola_2001}, provides an intuitive description of this transformation. The Jones matrix of the BM can be expanded as
\begin{equation}
\mathbb{J}(\bx) = q_0\sigma_0 + \bq\cdot\bm{\sigma},
\end{equation}
where $\sigma_0$ is the $2\times 2$ identity matrix and $\bm{\sigma}=(\sigma_1,\sigma_2,\sigma_3)$, with
\begin{equation}
\sigma_1= \JonesMat{\, 0}{-\ui}{-\ui}{\, 0},\quad
\sigma_2= \JonesMat{0}{-1}{1}{\, 0},\quad
\sigma_3= \JonesMat{-\ui}{0}{\, 0}{\ui}.
\end{equation}
Then $\mathbb{J}(\bx)$ can be regarded as a quaternion with basic units $\sigma_n$ $(n=1,2,3)$ since
$\sigma_1^2 \!=\! \sigma_2^2 \!=\! \sigma_3^2 \!=\! \sigma_1\sigma_2\sigma_3 \!=\! -\sigma_0$.
To describe the transformation of the input Stokes parameters by the BM, we use the polarization matrix \cite{Wolf_1954}
\begin{equation}
\mathbb{W} = \langle{\bf E}_0{\bf E}_0^\dagger\rangle_t = \tfrac{1}{2}S_0(\sigma_0 + \ui\sp\bs\cdot\bm{\sigma}),\label{eq:W_quat}
\end{equation}
where $\langle\cdot\rangle_t$ indicates a temporal average in the case of partially polarized light, and the Stokes parameter $S_0$ is the total intensity.
The polarization matrix after the BM is then $\mathbb{W}'=\mathbb{J}\mathbb{W}\mathbb{J}^\dagger$. Since $\mathbb{J}^\dagger=\mathbb{J}^{-1}$, this leads to the relations $S_0'=S_0$ and
\begin{equation}
\bs'\cdot\bm{\sigma} = \mathbb{J}\sp(\bs\cdot\bm{\sigma})\mathbb{J}^{-1},\label{eq:stokes_quat_rotation}
\end{equation}
where $S_0'$ and $\bs'$ are the output Stokes parameters. The pure quaternions $\bs\nsp\cdot\nsp\bm{\sigma}$ and $\bs'\nsp\cdot\nsp\bm{\sigma}$ correspond to points in a three-dimensional space (namely the Poincar\'e sphere), so Eq.~(\ref{eq:stokes_quat_rotation}) describes a rotation by the unit quaternion $\mathbb{J}$. By explicitly writing
\begin{equation}
\mathbb{J}(\bx) = \sigma_0 \cos\delta + \hat{\bq}\cdot\bm{\sigma}\sin\delta,
\end{equation}
one can see that indeed the axis of rotation is the unit vector $\hat{\bq}=\bq/\abs{\bq}$ and the angle of rotation is equal to the retardance $2\delta$, following the right-hand rule as shown in Fig.~\ref{fig:PoincareRotation}.

\begin{figure}[ht]
	\centering
	\includegraphics[width=.8\linewidth]{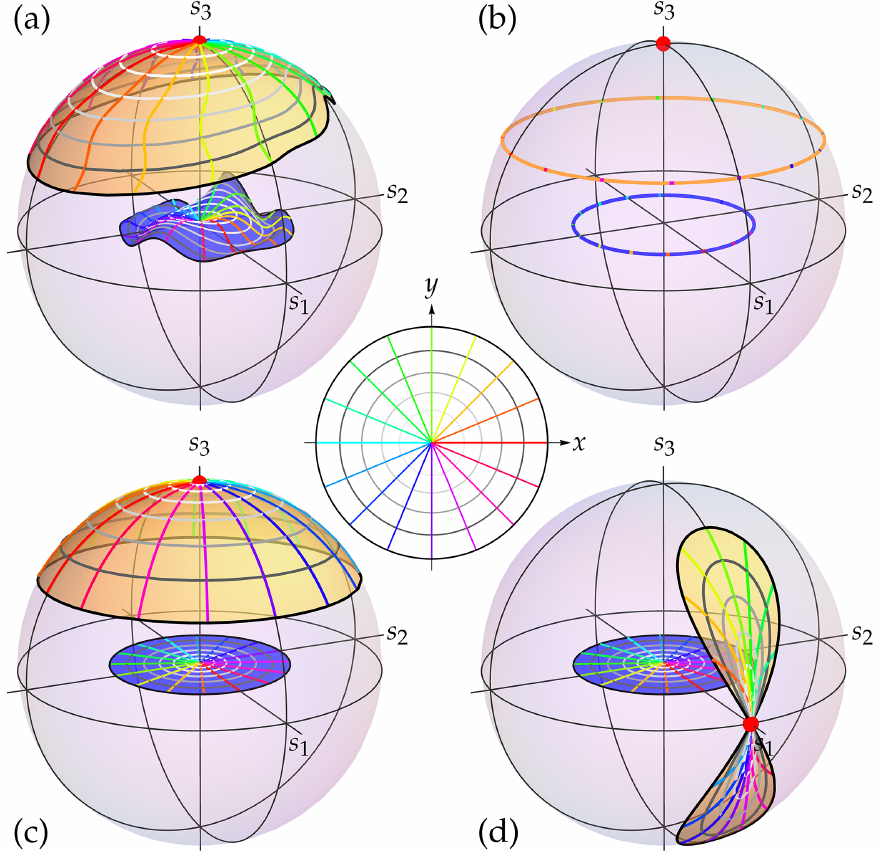}
	\caption{Input (red) and output (orange) polarizations of various BM distributions (blue). \VellaChange{Each surface is mapped to the spatial coordinate $\bx$ according to the radial (grayscale) and azimuthal (colored) contours shown at center.} The BM distributions are (a) an irregular BM, (b) a $q$-plate with retardance $\pi/3$, and (c-d) an SEO with maximum retardance $\pi/3$. The input polarizations are (a-c) right-circular and (d) horizontal.}
	\label{fig:PoincareBM}
\end{figure}

Taking into account the spatial variation of the BM, the distribution $\vec{q}(\bx)$ corresponds to a surface on the Poincar\'e hypersphere, which can be projected onto a surface $\bq(\bx)$ within the solid Poincar\'e sphere. Therefore, a uniformly polarized incident field is transformed into a spatially varying polarization, which also occupies a surface on the Poincar\'e sphere, as seen in Fig.~\ref{fig:PoincareBM}. \VellaChange{The irregular distribution shown in Fig.~\hbox{\hyperref[fig:PoincareBM]{\ref{fig:PoincareBM}(a)}} serves to illustrate the general case of a BM with elliptical eigenpolarizations. For devices with linear eigenpolarizations, such as the $q$-plate and stress-engineered optic (SEO) \hbox{\cite{Spilman_2007_vortex}} shown in Figs.~\hbox{\hyperref[fig:PoincareBM]{\ref{fig:PoincareBM}(b-d)}}, $\bq(\bx)$ is confined to the equatorial disk.} Notice that a $q$-plate and an SEO convert uniform right-circular polarization into distributions occupying a ring and a spherical cap, respectively. In the $\delta=\pi/2$ limit, the $q$-plate produces a left-circularly polarized beam with a phase vortex having the same topological charge as its eigenpolarization pattern \cite{Marrucci_2006}. Similarly, if the stress coefficient of the SEO is increased until $\bq(\bx)$ spans the equatorial disk, then any incident polarization is transformed into a beam which covers the entire surface of the sphere, with a particularly simple polarization mapping occurring for the case of circularly polarized input \cite{Beckley_2010}.
Using the quaternion representation, one can also see that if uniform waveplates are inserted on either side of a BM, the surface $\vec{q}(\bx)$ undergoes a rigid rotation in four dimensions \cite{Thomas_2016}, which will in general alter the shape of its projection $\bq(\bx)$. 
However, the effect of a uniform waveplate after the BM is simply a rigid rotation of $\bs'(\bx)$ in three dimensions.%

\section{Geometric phase} 
\begin{figure}[h]
	\centering
	\includegraphics[width=.8\linewidth]{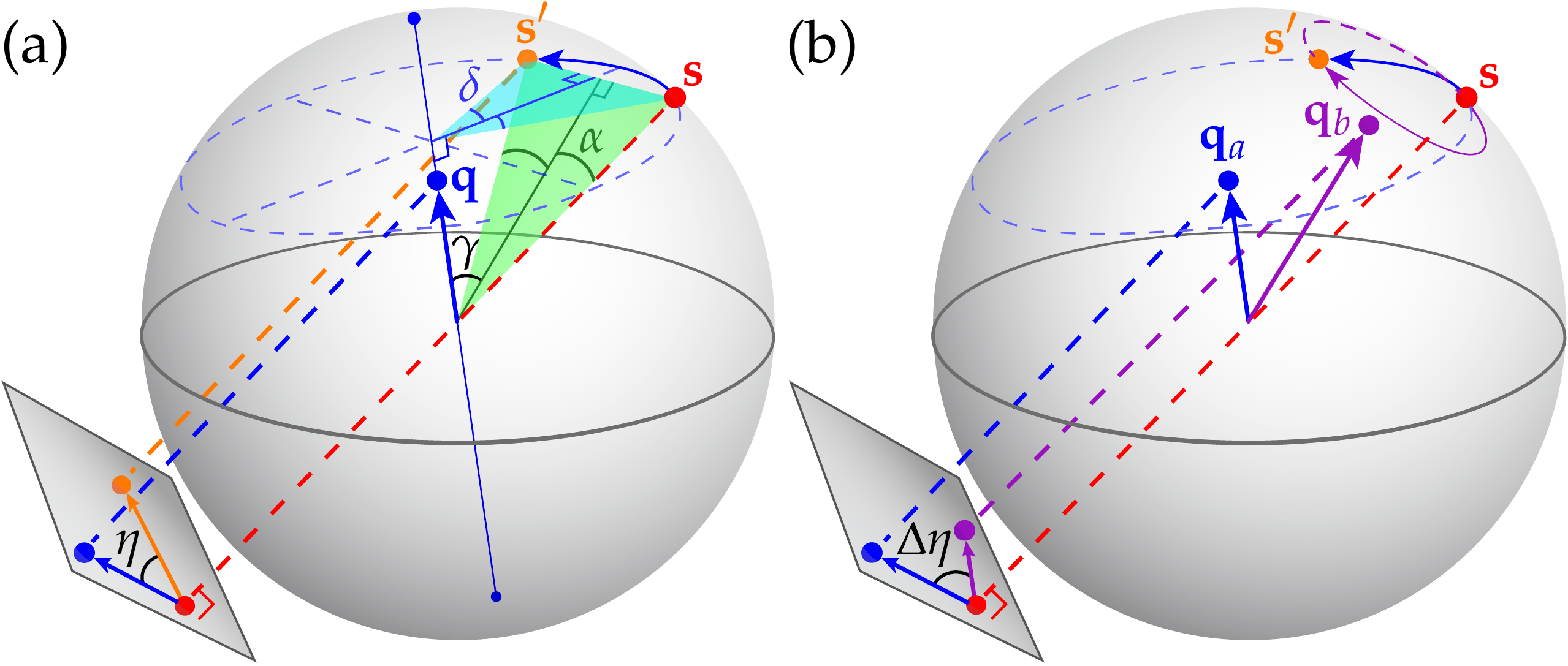}
	\caption{Visualization of (a) the geometric phase $\eta$ due to a birefringence vector $\bq$ and (b) the geometric phase difference $\Delta\eta$ between two transformations by vectors $\bq_a$ and $\bq_b$.}%
	\label{fig:geophase}%
\end{figure}

This representation provides not only a geometric description of the transformation of polarization but also of the associated geometric (Pancharatnam-Berry) phase \cite{Pancharatnam_1956,Berry_1987} of the resulting field at each point. Suppose for example that a uniform input polarization is transformed into a given output polarization by two different points of the BM. The difference in phase between the field at these two points of equal polarization is not necessarily proportional to the solid angle enclosed by the two circular trajectories over the Poincar\'e sphere, since these trajectories are in general not geodesic \cite{Bhandari_1989}. To understand this phase geometrically, consider an input polarization ${\bf e}$ represented by a point ${\bf s}$ on the Poincar\'e sphere, which is transformed by a BM into an output polarization ${\bf e}'$ represented by a Poincar\'e sphere point ${\bf s}'$. There are many Jones matrices that could achieve this transformation, because there are many ways to rotate the sphere so that ${\bf s}$ becomes ${\bf s}'$. However, it is easy to see that, by symmetry, the axis of rotation (the direction of ${\bf q}$) must be contained within the plane that bisects ${\bf s}$ and ${\bf s}'$. Further, as can be seen from Fig.~\ref{fig:geophase}(a), the angle of rotation (or retardance) $2\delta$ is related to the angle between ${\bf s}$ and ${\bf s}'$, referred to here as $2\alpha$, through the simple relation $\tan\delta=\tan\alpha/\sin\gamma$, where $\gamma$ is the angle between ${\bf q}$ and the bisector of ${\bf s}$ and ${\bf s}'$. Using this result, one can find a continuum set of vectors ${\bf q}(\gamma)$, and therefore of Jones matrices $\mathbb{J}(\gamma)$, that achieve the desired transformation. A tedious but straightforward calculation shows that $\mathbb{J}(\gamma)\cdot{\bf e}\propto\exp[\ui(\Gamma+\eta)]{\bf e}'$,
where $\tan\eta=\tan\gamma/\sin\alpha$. Here $\Gamma$ and $\eta$ are the dynamic and geometric phases, respectively, imparted on the input field by the BM. Therefore, the geometric phase difference between two BM transformations that take a given input polarization to a given output polarization equals the difference in their phases $\eta$. From Fig.~\ref{fig:geophase}(a) one can see that $\eta$ has a simple geometric interpretation. Suppose that the Poincar\'e sphere is projected onto a plane normal to the input polarization ${\bf s}$. Then $\eta$ is the angle between the projections of ${\bf s}'$ and ${\bf q}$. From this interpretation, one can infer that the geometric phase difference resulting from two different birefringence matrices that achieve the same final polarization from the same initial one is equal to the angle $\Delta\eta$ between the projections of their vectors ${\bf q}$ onto the plane perpendicular to the Stokes vector of the input polarization, as shown in Fig.~\ref{fig:geophase}(b).
\FloatBarrier

\section{Effect on imaging systems}\label{sect:PSF_overview}
As an example, we now apply this formalism to optical systems that include a focusing element, and where the BM is at a Fourier-conjugate plane to the final plane where the intensity is measured. The most common scenario is that of an (exit-telecentric) imaging system, where the BM is assumed to be placed at the pupil plane. The position-dependent retardance could be caused by undesired stress in the optical elements, or by the intentional inclusion of a BM, e.g., for polarimetric applications \cite{Ramkhalawon_2013, Zimmerman_2016, Sivankutty_2016}. (If the birefringence is due to elements not at the pupil plane, an approximation from aberration theory can be used in which the errors are accumulated at the pupil by transporting them there along the system's nominal rays \cite{Youngworth_2000}.) In what follows, we study the effect of a BM at the pupil plane on the point-spread function (PSF). These results are then used to quantify the effects on measures of image quality, such as the size of the PSF and the Strehl ratio.

\begin{figure}
	\centering
	\includegraphics[width=.8\linewidth]{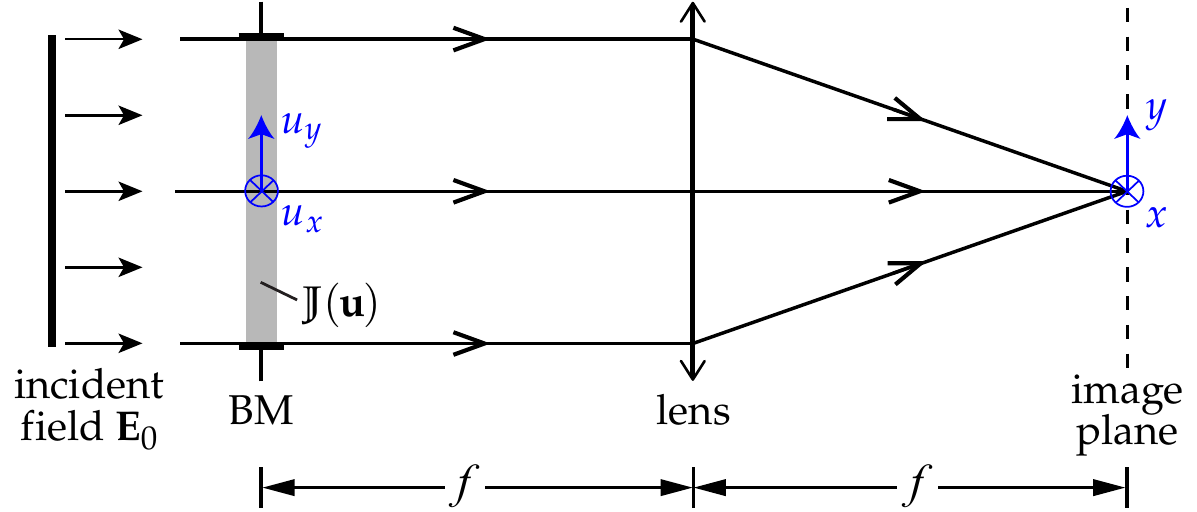}
	\caption{Exit-telecentric imaging system with a BM and aperture at the front focal plane of a lens with focal length $f$.}
	\label{fig:BM_focusing_layout}
\end{figure}

Suppose that a uniformly polarized input field ${\bf E}_0$ is incident on a BM at the plane of an aperture with real pupil function $A$ (binary or apodized) and is focused by a lens, as shown in Fig.~\ref{fig:BM_focusing_layout}. 
Let $\bu$ and $\bx$ represent the spatial coordinates in the pupil and image planes, respectively, where $\bu$ is normalized such that in the paraxial limit, its magnitude is equal to the focusing angle after the lens.
The field ${\bf E}(\bx)$ in the image plane is related to the pupil distribution via the Fourier transformation
\begin{equation}
{\bf E}(\bx)= \mathcal{F}\{A\sp\mathbb{J}\cdot{\bf E}_0\} =
\int \!A({\bf u})\,{\mathbb J}({\bf u})\cdot{\bf E}_0\exp\left[\ui k({\bf u}\cdot\bx)\right]\ud^2u. \label{eq:E(x)}
\end{equation}
The PSF $I(\bx)=\langle\abs{{\bf E}(\bx)}^2\rangle_t
$ may then be written as
\begin{align}
I(\bx) = \iint &A(\bu_1)A(\bu_2) \sp{\rm Tr}\nsp\left[\mathbb{J}(\bu_2)\cdot\mathbb{W}\cdot\mathbb{J}^\dagger(\bu_1)\right]\nonumber\\[-1pt]
&\times\exp[\ui k(\bu_2-\bu_1)\cdot \bx]\ud^2u_1\ud^2 u_2.\label{eq:I(x)_W}
\end{align}
For the common case of an unpolarized input field ($\mathbb{W}=\sigma_0$), this reduces to
\begin{equation}
I(\bx)=2\abs*{\mathcal{F}\{A\exp(\ui\Gamma)\vec{q}\}}^2.\label{eq:I(x)_Fourier}
\end{equation}%

The RMS width $r$ of the PSF (with respect to the ideal focal point $\bx={\bf 0}$) is given by
\begin{equation}
r^2=\frac{\int \abs*{\bx}^2I(\bx)\,\ud^2x}{\int I(\bx)\,\ud^2x}=\frac{1}{k^2}\frac{\int \norm*{\nabla[A\exp(\ui\Gamma)\vec{q}]}^2\,\ud^2u}{\int \abs{A\exp(\ui\Gamma)\vec{q}}^2\,\ud^2u},
\label{eq:r2}
\end{equation}
where in the second step we used Parseval's theorem and the Fourier property $\bx\to-(\ui/k)\,\nabla$, with $\nabla$ being the gradient in $\bu$. The integrand in the numerator is the squared Frobenius norm of the $4\times 2$ Jacobian matrix of derivatives of $A\exp(\ui\Gamma)q_n$ ($n=0,1,2,3$) with respect to the components of $\bu$. One can expand this expression to obtain the surprisingly simple result
\begin{equation}
r^2=\frac{1}{k^2}
\frac{\int 
	(\abs{\nabla A}^2+A^2\abs{\nabla\Gamma}^2 + A^2\norm{\nabla\vec{q}}^2 )\ud^2u}
{\int\! A^2\ud^2u}. \label{eq:r2_expanded}
\end{equation}
%
Each of the three terms within the parentheses in the numerator has an intuitive interpretation as a contribution to the PSF's RMS width. The first accounts for diffraction by the aperture. Note that if $A$ represents a hard aperture, the RMS width is not well-defined since this term diverges; it is well-defined only if the pupil function $A$ represents an apodized pupil that provides a continuous transition to zero. The second term accounts for the effects of variations of the global phase $\Gamma$, which can be regarded as a standard phase aberration. The third term is the one that accounts for variations in birefringence. Here, the Frobenius norm $\norm*{\nabla\vec{q}}$ indicates the rate of change of $\vec{q}$ over the Poincar\'e hypersphere as the pupil position varies. Notice that the second and third terms can also be interpreted respectively as the effects of dynamic and geometric phase discrepancies over the pupil.

Next we consider the effect of the BM on the Strehl ratio, defined as the intensity at the center of the PSF, normalized by the same quantity when there is no birefringence or aberration:
\begin{equation}
\mathcal{S} = \frac{I({\bf 0})}{\bigl.I({\bf 0})\bigr|_{\delta=\Gamma=0}}
=\frac{\abs*{\int\! A\exp(\ui\Gamma)\vec{q}\,\ud^2 u}^2}{\left(\int\! A\sp\ud^2 u\right)^2},
\label{eq:strehl}
\end{equation}
where in the second step we again assumed an unpolarized input. In the absence of aberrations (constant $\Gamma$), $\smash{\mathcal{S}=\abs{\langle\vec{q}\rangle_A}^2}$, where $\langle\vec{q}\rangle_A$ is the spatial average of $\vec{q}$ over the pupil, weighted by the pupil function $A$. That is, if one densely samples the pupil uniformly and finds all points $\vec{q}(\bu)$, the distance from the origin to their centroid on the Poincar\'e hypersphere is the square root of the Strehl ratio. Since $\vec{q}(\bu)$ is constrained to the surface of the Poincar\'e hypersphere, it follows that ${\mathcal{S}\leq 1}$, with the equality occurring when $\vec{q}$ is constant. Note that Eqs.~(\ref{eq:r2_expanded}) and (\ref{eq:strehl}) are explicitly invariant under global unitary transformations caused by the cascading of uniform waveplates, since these only cause a rotation of the surface $\vec{q}(\bu)$ over the hypersphere.

\FloatBarrier
\section{Concluding remarks}
We proposed a graphic representation for visualizing the effects of spatially varying birefringence on an incident field, including both polarization and geometric phase. The basis for this representation is the definition of a Poincar\'e hypersphere, and then its ``vertical'' projection onto the Poincar\'e sphere by dropping one component, $q_0$ (except for its sign), from the four-dimensional space. Note that other projections of the 3D hypersurface of half the unit Poincar\'e hypersphere onto a flat 3D region could have been used, such as the ``central'' projection ${\bf q}/q_0$, which naturally accounts for the effects of a sign change in $q_0$, and for which the length of the vector is $\tan\delta$ instead of $\sin\delta$. However, the resulting distribution would occupy all space, while the mapping used here occupies only a solid unit sphere. The convenience of being restricted to a finite space that is also occupied by all possible polarization states of the field is the reason for our choice of mapping. 

For imaging or focusing systems that include birefringence in their pupil plane, the distribution of $\vec{q}$ over the Poincar\'e hypersphere enters naturally into the image quality metrics, its effects being particularly simple if the object is unpolarized: there is an increase in the RMS area of the PSF that is directly related to the rate of change of $\vec{q}$ over the pupil, while the Strehl ratio is the squared magnitude of the centroid vector of the distribution of $\vec{q}$ over the Poincar\'e hypersphere for all points in the pupil. 
These results are also relevant to 
non-imaging applications that use similar configurations. For example, the birefringence distribution in the pupil plane can be tailored such that the focused field forms an optical bottle ($\mathcal{S}=0$) for circular input polarization \cite{Vella_2017_bottle,Vella_2017_bottle_erratum}. The formalism introduced here can also be applied to optimize the polarimetry technique used in \cite{Ramkhalawon_2013,Zimmerman_2016, Sivankutty_2016}. The results will be discussed in a future publication.


\section*{Acknowledgments}
The authors would like to thank Konstantin Bliokh and Thomas G. Brown for useful comments. This work was funded by the National Science Foundation (NSF) (PHY-1507278).

\end{document}